\documentclass[pre,aps,draft,showpacs,preprint,superscriptaddress]{revtex4}
\usepackage[dvips,final]{graphicx}
\usepackage{bm}

\begin{document}

\title{External and intrinsic anchoring in nematic liquid crystals: A Monte Carlo study}
\author{N. V. Priezjev}  
\affiliation{Department of Physics, Brown University, Providence, Rhode Island 02912}
\author{G. Ska\v cej}
\affiliation{Department of Physics, University of Ljubljana, Jadranska 19,
             SI-1000 Ljubljana, Slovenia}
\author{R. A. Pelcovits}
\affiliation{Department of Physics, Brown University, Providence, Rhode Island 02912}
\author{S. \v Zumer}
\affiliation{Department of Physics, University of Ljubljana, Jadranska 19,
             SI-1000 Ljubljana, Slovenia}

\date{\today}

\begin{abstract}

We present a Monte Carlo study of external surface anchoring in nematic cells with partially
disordered solid substrates, as well as of intrinsic anchoring at free nematic interfaces.
The simulations are based on the simple hexagonal lattice model with a spatially anisotropic
intermolecular potential. We estimate the corresponding extrapolation length $b$ by
imposing an elastic deformation in a hybrid cell-like nematic sample. Our estimates for
$b$ increase with increasing surface disorder and are essentially temperature--independent.
Experimental values of $b$ are approached only when both the coupling of nematic molecules
with the substrate and the anisotropy of nematic--nematic interactions are weak.

\end{abstract}
\maketitle

\section{Introduction}

In confined nematic liquid crystals with a large surface--to--volume ratio the aligning
effects of the confining surfaces are of great importance in determining the equilibrium
director configuration~\cite{Zumer}. There are two major contributions to these surface
aligning effects, the first one originating from direct interactions between nematic
molecules and the solid substrate (external anchoring), while the second one is due
to incomplete anisotropic nematic-nematic interactions in the vicinity of the sample surface
(intrinsic anchoring). An understanding of these confinement related aligning mechanisms
is of great importance not only from the fundamental point of view, but also, e.g.,
for the design and construction of liquid crystal--based optical devices.
Within phenomenological approaches anchoring effects are usually characterized by two
parameters: the preferred molecular alignment direction at the sample surface (the
easy axis) and the free energy coefficient $W$ penalizing any deviation from this direction
(the anchoring strength)~\cite{Rapini}. Here the $W$ coefficient depends on $S$, the
nematic order parameter. This dependence seems to be strongly related to the specific
properties of a given interface: for example, for a system of hard rods confined between
hard walls one finds $W\propto S$~\cite{Okano}, while in experiments measuring
anchoring inside polycarbonate membranes even~$W\propto S^4$ can be obtained~\cite{Alenka}.
Given, moreover, $K\propto S^2$~\cite{deGennes,Cleaver} ($K$ denoting the Frank elastic
constant), with decreasing $S$ the extrapolation length $b=K/W$~\cite{deGennes} may
either increase (as seen experimentally in thermotropics~\cite{Alenka}), or decrease
(as obtained from simulations with hard particles~\cite{Andrienko}). In the bulk, the
value of $S$ is primarily determined by temperature, while close to an interface it may
also be affected by incomplete bulk interactions, as well as by (dis)ordering effects of
the (possibly rough) confining substrate. In the latter case $W$ is often assumed to be
simply proportional to ${\cal W}S_0S(0)$, where ${\cal W}$ represents the surface coupling
constant, $S_0$ the surface--imposed value of $S$ (determined by substrate roughness), and
$S(0)$ its actual surface value~\cite{Nobili92}.

While both the anchoring strength and the easy axis can be determined
experimentally~\cite{Blinov}, it is also possible to deduce them
from simulations based on pairwise intermolecular interactions. For example, intrinsic
anchoring has been analyzed in a pseudomolecular continuum approach with ellipsoidal
molecules~\cite{Silvia}, using a lattice approximation in the zero--temperature
limit~\cite{hex}, or analyzing different types of interfaces in Gay--Berne
systems~\cite{Emerson,Mills,Bates97}. Furthermore, surface anchoring strength
has also been measured in a system of hard ellipsoids in contact with a hard
wall~\cite{Andrienko,Allen}. The anchoring strength reported in most analyses
shows that anchoring is rather strong and that the corresponding extrapolation
length is of the order of a few molecular dimensions, while its experimental values are
typically above 100~nm~\cite{Blinov}. On the other hand, a recent analysis of external
anchoring --- also based on a lattice model, but for nonzero temperature --- was presented
in Ref.~\cite{Nikolai}, showing that the extrapolation length can increase significantly
when the nematic--to--isotropic (NI) transition is approached, as also seen
experimentally~\cite{Alenka}. Moreover, a lattice gas approach has been adapted
recently to study nematic interfaces~\cite{Bates02}.

Motivated by these developments, in this paper we extend the analyses performed in
Refs.~\cite{hex,Nikolai} to nonzero temperatures (employing Monte Carlo simulations)
and repeat the measurement of the extrapolation length both for external and intrinsic
anchoring, the former in presence of rough solid substrates and the latter for a free
nematic interface. The term ``rough'' here refers to flat but partially disordered
substrates. We will first briefly recall the features of the lattice model used in
Ref.~\cite{hex}, then discuss the modifications needed to perform the present
analysis, and finally, present and discuss the results.

\section{Simulation model and anchoring measurement}

We model a liquid crystalline slab of thickness $d$, using a modification
of the well--known Lebwohl--Lasher (LL) model~\cite{Lebwohl}. In the LL model
elongated nematic molecules are represented by freely-rotating spin-like
particles that are attached to lattice points of a cubic lattice. Since the
pairwise interaction energy used in the LL model is spatially isotropic and
thus does not depend on the relative position of the particles, it cannot produce any orienting
effects at a free nematic surface and is therefore not suitable for studies of
intrinsic anchoring. Therefore, we choose a more general potential to model the interparticle
interaction energy resulting from anisotropic van der Waals forces. These forces can
give rise to the nematic phase~\cite{Maier} even without hard rod repulsion. For
two neighboring nematic particles $i$ and $j$ with orientations given by unit vectors
${\bf u}_i$ and ${\bf u}_j$, separated by ${\bf r}_{ij}$, we model the interaction potential
as~\cite{Nehring}:
\begin{eqnarray}
 U_{ij}=-\epsilon\left[
        {\bf u}_i\cdot{\bf u}_j - 3\nu({\bf u}_i\cdot{{\bf r}_{ij}\over r_{ij}})
                                      ({\bf u}_j\cdot{{\bf r}_{ij}\over r_{ij}})\right]^2  \label{1}
\end{eqnarray}
with $\epsilon>0$, which for $\nu=0$ reduces to the standard LL potential, while
for $\nu=1$ one has the induced dipole-induced dipole interaction. The spatial
anisotropy parameter $\nu$ enables us to continuously vary the relative
importance of the spatially anisotropic (${\bf r}_{ij}$-dependent) contribution to the
interaction law~(\ref{1}). Although the range of van der Waals potential is
proportional to $r_{ij}^{-6}$, for computational efficiency we consider only interactions
between nearest neighbors. Thereby the intrinsic anchoring energy $W$ and the elastic
constant $K$ are underestimated, but we expect the errors in the estimation of $W$ and
of the extrapolation length $K/W$ not to exceed 20\% and 30\%, respectively.

In a nematic slab, boundary conditions are typically fixed by the interaction with solid
walls, or, alternatively, through orienting effects near free nematic interfaces. In
the simulation, each of the solid walls is represented by a layer of fixed particles
${\mathbf p}_i$ ($|{\mathbf p}_i|=1$), either all perfectly aligned or somewhat
disordered with some residual order. The nematic--wall interaction is modeled via
\begin{eqnarray}
 U_{ij}^s=-\epsilon_s\left[{\mathbf p}_i\cdot{\mathbf u}_j\right]^2,  \label{2}
\end{eqnarray}
as promoted, e.g., by short--range steric forces, where, again, ${\mathbf p}_i$ and ${\mathbf u}_j$
are nearest neighbors on the lattice. A dimensionless anchoring strength parameter
can be defined as $w=\epsilon_s/\epsilon$. In the case of a perfectly aligned surface
one has ${\mathbf p}_i={\mathbf\Pi}$, i.e., all particles are aligned along the easy axis
${\mathbf\Pi}$, while for disordered surfaces ${\mathbf\Pi}$ represents the average orientation
of ${\mathbf p}_i$. On the other hand, a free nematic interface is simply modeled through
the missing--neighbor effect.

\begin{figure}
 \begin{center}
 \includegraphics[width=86mm]{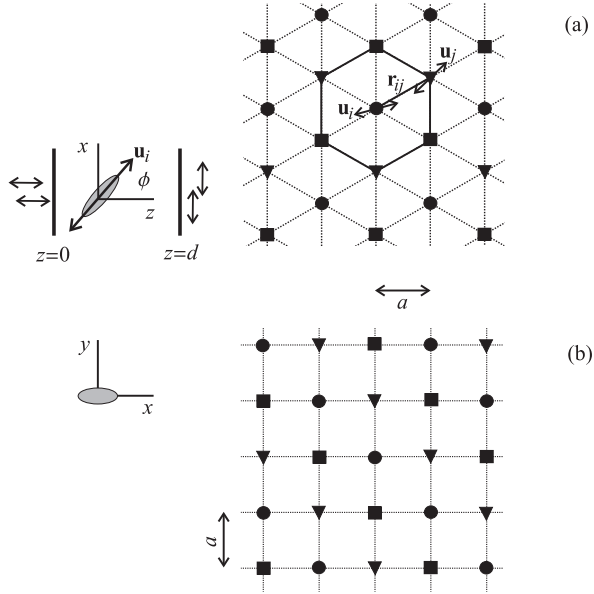}
 \caption[]{Sample geometry, hexagonal lattice and the three sublattices (squares,
            circles, and triangles): (a)~$xz$ and (b)~$xy$ cross sections. The tilt
            angle $\phi=\phi(z)$ is measured with respect to the $z$-axis, the sample
            normal. Boundary conditions at $z=0$ and $z=d$ are homeotropic and planar,
            respectively.}
 \label{sl1}
 \end{center}
\end{figure}

Instead of the cubic lattice used in the LL model, in present simulations we use
a simple hexagonal lattice to model the liquid crystal, which is necessary to
avoid unphysical bulk easy axes~\cite{hex} for $\nu\neq 0$. These axes arise as a direct
consequence of using a lattice approximation for modeling an anisotropic liquid
and are present for any spatially anisotropic potential in the cubic lattice.
They are absent, however, in the hexagonal lattice, provided that the spins
${\mathbf u}_i$ are assumed to be two--dimensional vectors confined to
hexagonal planes. The geometry of our sample is shown in Fig.~\ref{sl1}. The
$z$ axis is normal to the confining interfaces and the hexagonal planes are parallel
to the $xz$ plane. We stress that our analysis must be restricted to low
values of the anisotropy parameter $\nu$ in order to avoid solid-like periodic
director profile solutions that are stable for $\nu>0.3$~\cite{hex}.

The strength of any anchoring can be measured by imposing an elastic
distortion so that the average surface molecular orientation deviates from the
easy axis defined by the anchoring. The magnitude of this deviation can then be
used to estimate the anchoring strength and the corresponding extrapolation
length~\cite{deGennes}. The elastic distortion in a nematic slab can be imposed
either by applying a magnetic field whose orientation must not coincide with the
direction of the easy axis, or by antagonistic anchoring conditions at opposing
surfaces. In the zero-temperature analysis of Ref.~\cite{hex} the magnetic field
method was used. On the other hand, for nonzero temperatures a strong enough magnetic
field can enhance the degree of nematic order and even shift the NI phase transition,
which can present additional difficulties in interpreting the results. Therefore, we
decided not to use the magnetic field approach; instead, we consider a hybrid cell-like
sample with antagonistic boundary conditions.

Consider the $\nu=0$ case with pure external anchoring first. The left ($z=0$)
interface --- where the anchoring will be measured --- is chosen to be a solid wall, either
perfectly ordered or somewhat disordered, promoting homeotropic alignment through weak external
anchoring ($w=0.1$). At the same time, the right ($z=d$) interface is also taken to be a
solid wall, yet with perfect planar alignment and strong external anchoring ($w=1$).
For the case $\nu\neq 0$, we replace the solid wall at $z=0$ with a free nematic
interface, enabling us to study intrinsic anchoring alone. Note that for $\nu\le 0.3$
the easy axis of the free interface is homeotropic~\cite{hex}, thus providing the same
confinement type as with external anchoring. As a result, in both cases a combined
bend and splay elastic deformation is expected to appear in the sample. The deformation
should be present as long as the sample thickness $d$ exceeds $d_c=\vert(K/W)_0-(K/W)_d\vert$,
where $(K/W)_0$ and $(K/W)_d$ are the extrapolation lengths corresponding to the effective
anchoring on the left and the right wall, respectively, while for $d<d_c$ a uniform director
structure is observed~\cite{Barbero}. In a hybrid cell close to NI transition a non-bent
biaxial structure is also possible, consisting of two strata of uniform alignment, separated
by a biaxially ordered layer~\cite{Andreja}. Recall that unlike in the original LL model
in the present study spins ${\mathbf u}_i$ are two--dimensional vectors and hence
unable to reproduce biaxiality.

For hybrid boundary conditions in the one--constant approximation the Frank
elastic theory (assuming a constant degree of nematic order throughout the slab)
predicts a perfectly linear director tilt angle profile $\phi(z)$, where $\phi$
is measured, e.g., with respect to the slab normal $z$. The torque balance condition
at, e.g., the left surface ($z=0$) can be written as $(d\phi/dz)_0={1\over 2}(W/K)_0\sin{2\phi(0)}$
and enables us to deduce $(K/W)_0$ from the measured values of $(d\phi/dz)_0$ and $\phi(0)$.
Note that for small $\phi(0)$ (i.e., strong enough anchoring) the above condition
simplifies to $(d\phi/dz)_0=(W/K)_0\phi(0)$ and allows us to determine $(K/W)_0$
simply by extrapolating the profile $\phi(z)$ graphically across the sample
boundary to $\phi=0$ corresponding to the homeotropic easy axis. Note also
that if the degree of nematic order is subject to variations --- which is usually the
case near sample boundaries in a layer of thickness $\xi$ ($\xi$ denoting
the nematic correlation length) --- the profile $\phi(z)$ may deviate from the
above predicted linear behavior. In this case the extrapolation of the profile
towards the surface must be performed from far enough in the bulk where the order
parameter profile is constant, that is from $\xi<z<d-\xi$.
Note also that whenever $(K/W)_0$ approaches $d$ (for weak anchoring or in a thin sample),
$(d\phi/dz)_0$ and, consequently, $(K/W)_0$ are accompanied by a significant systematic error.

\begin{figure}
 \begin{center}
 \includegraphics[width=75mm]{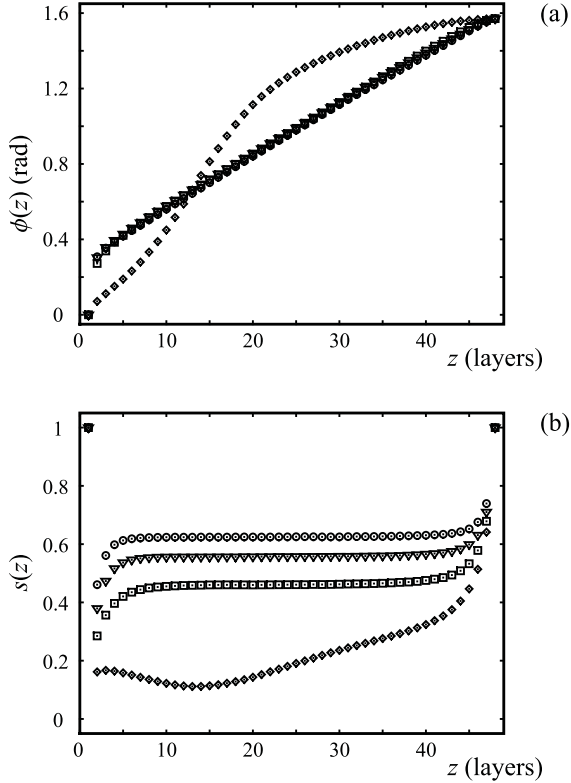}
 \caption[]{Hybrid cell with smooth substrates and $\nu=0$ (external anchoring at $z=0$):
            director~(a) and order parameter~(b) profiles for $T^*=1.50$, $1.42$, $1.34$,
            and $1.26$ (diamonds, squares, triangles, and circles, respectively); the bulk
            NI transition temperature approaches $T_{NI}^*\approx 1.52$.}
 \label{sl2}
 \end{center}
\end{figure}

Our Monte Carlo (MC) simulations are now performed as follows. The size of the simulation
box size was set to $48^3$ for a total of 105984 nematic particles, excluding the boundary
particles ${\mathbf p}_i$ in the layers at $z=0$ and $z=d$. In case of disordered
substrates particle orientations in the confining layers are generated following a
probability distribution $f(\phi)\propto\exp{(-{\cal P}\cos^2{\phi})}$, which gives
a homeotropic easy axis ($\phi=0$) and a ${\cal P}$-dependent degree of order $S_0$.
For example, ${\cal P}\to 0$ corresponds to a distribution close to isotropic, while
${\cal P}\to\infty$ yields a perfectly aligned substrate. Then, the hexagonal lattice is
divided into three sublattices, as shown in Fig.~\ref{sl1}, ensuring that the bonds
between neighboring particles on the lattice never connect two particles from the
same sublattice. Considering the simple hexagonal lattice as tripartite enables us
to vectorize the simulation algorithm, which provides a significant speed-up in
calculations. Further, in the $x$ and $y$ directions periodic boundary conditions
are assumed. We start either from a random configuration in two dimensions (recall
that ${\mathbf u}_i$ are restricted to lie in hexagonal planes), or from an
equilibrated configuration at a temperature slightly higher than the simulated
one, if this is available. Then we apply the standard Metropolis
algorithm~\cite{Metropolis}. For our vectorized algorithm to work
correctly, in each MC cycle we first attempt (and accept/reject) trial moves
involving particles in the first sublattice and only then proceed to the
second one, and after that to the third one. In generating a new trial
configuration each time only a single particle is involved. We have typically
performed $10^5$ MC cycles for equilibration, followed by
$10^5$ production cycles to accumulate averages of interest.

We measure the extrapolation length by analyzing the director profile
$\phi(z)$. The $\phi(z)$ dependence is calculated by accumulating the two
independent components of the two--dimensional ordering matrix $Q_{\alpha\beta}(z)=
2\langle u_i^\alpha u_i^\beta\rangle_z-\delta_{\alpha\beta}$, where $\alpha$ and
$\beta$ can be either $x$ or $z$, and $u_i^\alpha$ represents the $\alpha$
component of the unit vector ${\mathbf u}_i$. The average $\langle...\rangle_z$
is performed both over all particles in the layer centered at $z$ and over
the production MC cycles. Then, the averaged ordering matrix is diagonalized
and the positive eigenvalue is identified as the two--dimensional scalar order
parameter $s(z)$. Accordingly, the corresponding eigenvector is the director.
Note that in Ref.~\cite{Nikolai} the extrapolation length in the LL model with
three--dimensional spins was studied, using a different method to obtain the
director profile. Rather than computing the ordering matrix, the polar angle
averaged over all particles in a layer and over production MC cycles was
computed; i.e., no information was kept about the azimuthal orientation of
the particles. This method overestimates the actual average director tilt
angle for small values of this angle. The strong temperature dependence of
the extrapolation length found in Ref.~\cite{Nikolai} is most likely an
artifact of the incorrect method used in that paper.

\section{External anchoring}

We analyze the spatially isotropic intermolecular interaction first by setting $\nu=0$
in Eq.~(\ref{1}). Such a model is now similar to the standard LL model,
yet it is characterized by a different coordination number (eight in the former model,
six in the latter) and the dimensionality of nematic spins (two in the former model,
three in the latter). These distinctions lead to a different balance between the
decrease of internal energy and loss of orientational entropy upon going from the
isotropic into the nematic phase, and thus shift the NI transition to a temperature
higher than in the LL model. Monitoring temperature scans of internal energy, specific
heat, and the order parameter $s$ (lowercase $s$ will be used throughout the text for
the two--dimensional order parameter), for $\nu=0$ we estimate the dimensionless
transition temperature in a $48^3$ bulk sample (with full periodic boundary
conditions) to be $T_{NI}^*=k_B T_{NI}/\epsilon\approx 1.52\pm 0.01$. Here
a dimensionless temperature scale, $T^*=k_B T/\epsilon$, has been introduced.

\begin{figure}
 \begin{center}
 \includegraphics[width=75mm]{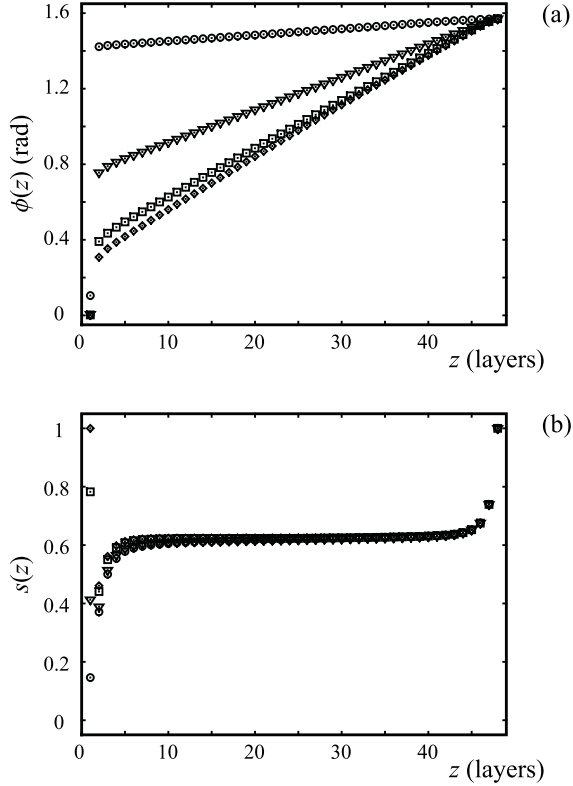}
 \caption[]{Hybrid cell with a rough homeotropic substrate at $z=0$: director~(a) and
            order parameter~(b) profiles for $\nu=0$, $T^*=1.26$, and different
            degrees of substrate roughness $s_0$. Circles, triangles,
            squares, and diamonds correspond to $s_0=1$, $s_0\approx 0.78$,
            $s_0\approx 0.41$, and $s_0\approx 0.15$, respectively.}
 \label{sl3}
 \end{center}
\end{figure}

Consider the director and order parameter profiles [$\phi(z)$ and $s(z)$, respectively]
in a hybrid nematic cell where both substrates are smooth and hence impose perfect
nematic order with $s_0=1$. Recall that for $\nu=0$ one is dealing exclusively
with external anchoring. Fig.~\ref{sl2}(a) shows $\phi(z)$ profiles for different
temperatures. One can readily observe that far enough from the NI transition the
profiles approach a linear function (predicted also from Frank elastic
theory~\cite{deGennes}), with minor changes in slope $|d\phi/dz|$ only close to
substrates where the degree of order can vary. In particular, wherever $s(z)$
exceeds its bulk value $s_b$, the nematic becomes more difficult to deform,
which results in a reduction in slope (and vice versa). Moreover, note that
far enough from the NI transition [in our simulations for values of reduced
temperature $\tau=(T_{NI}^*-T^*)/T_{NI}^*\gtrsim 0.07$] the $\phi(z)$ profiles
are essentially insensitive to changing temperature. Given $w=0.1$, as chosen
above for the homeotropically anchored smooth left wall, the extrapolation
length is estimated as $b=11.6a(1\pm 7\%)$, where $a$ is the lattice spacing.
Significant deviations from the linear $\phi(z)$ profile can be observed only
close to $T_{NI}^*$ [for $\tau\lesssim 0.02$], when the nematic far enough
from the walls melts and thereby avoids elastic distortion
--- see Fig.~\ref{sl2}~(b). Then molecular alignment becomes homeotropic in the
vicinity of the left surface, followed by a region of (nearly) isotropic liquid
in the slab center, and by a region of planar alignment close to the right surface.
Note that one can avoid this nematic ``meltdown'' by reducing the deformation strength,
e.g., by setting $0<\phi(d)\ll\pi/2$ or by reducing $w$ at the $z=d$ wall. In this case
$b$ is observed to be essentially $T^*$-independent even up to $\tau\approx 0.02$.
Note, however, that thereby the immediate vicinity of the NI transition actually
has not been probed: for a realistic liquid crystal with $T_{NI}\approx 300$~K,
$\tau\approx 0.02$ corresponds to temperatures as much as 6~K below the transition.
Alternatively, the deformation strength can be reduced also by significantly increasing
the system size. Further, in Fig.~\ref{sl2}~(b) one can always observe an increase of $s$ when
approaching the substrate at $z=d$ promoting planar alignment. Similarly,
$s$ decreases close to the $z=0$ substrate because of a weaker nematic--substrate
coupling ($w=0.1$ as opposed to $w=1$ for the $z=d$ substrate).

\begin{table}
 \caption{Extrapolation lengths $b$ measured in units of lattice spacing $a$:
          substrate roughness-dependence for external anchoring ($\nu=0$) and
          values for intrinsic anchoring ($\nu\neq 0$), together with the
          corresponding dimensionless bulk NI phase transition temperatures $T^*_{NI}$.
          The values are given for  $T^*\lesssim 1.42$ ($\nu=0$), $T^*\lesssim 1.265$
          ($\nu=0.05$), and $T^*\lesssim 1.13$ ($\nu=0.1$), i.e., for $\tau\gtrsim 0.07$,
          where $b$ is essentially temperature-independent.}
 \vspace*{5mm}
 \begin{ruledtabular}
 \begin{tabular}{r r r r r r}
    $\nu$ & $T^*_{NI}$      & anchoring  & $s_0$ & $b~(a)$  \\ \hline
     0    & $1.52 \pm 0.01$ & external   & 1.0  & $ 11.6\,(1\pm 7\%)$  \\
          &                 & & 0.78 & $ 14.9\,(1\pm 7\%)$  \\
          &                 & & 0.41 & $ 31\,(1\pm 7\%)$    \\
          &                 & & 0.15 & $ 50\,(1\pm 48\%)$   \\ \hline
     0.05 & $1.36 \pm 0.01$ & intrinsic & & $ 16.3\,(1\pm 13\%)$ \\ \hline
     0.1  & $1.21 \pm 0.01$ & intrinsic & & $ 4.6\,(1\pm 10\%)$  \\
 \end{tabular}
 \end{ruledtabular}
 \label{tabela}
\end{table}

Turn now to cases with a rough homeotropic substrate, simulated by generating an ensemble
of fixed particles ${\mathbf p}_i$ at $z=0$ with $s_0<1$. Qualitatively, the behavior of
$\phi(z)$ and $s(z)$ does not change, including the insensitivity of $b$ to variations of
$T^*$ in the temperature range presently accessible. However, as shown in Fig.~\ref{sl3}
for $T^*=1.26$, a smaller $s_0$ reflects in a reduced anchoring strength at $z=0$ and
hence in a reduced slope $|d\phi/dz|$, which, in turn, results in a larger $b$: for
$s_0\approx 0.78$ we find $b=14.9a(1\pm 7\%)$, for $s_0\approx 0.41$, $b=31a(1\pm 7\%)$,
and for $s_0\approx 0.15$, $b=50a(1\pm 48\%)$ (for a summary see Table~\ref{tabela}).
Note that in the latter case $b$ already approaches the sample thickness $d$ and is
already close to violating the stability condition $d\gtrsim b$ for the deformed
director structure~\cite{Barbero}. Moreover, assuming $K\propto s_b^2$ and
$W\propto s_0 s(0)$, the extrapolation length should scale as $b\propto s_b^2/s(0)s_0$.
The full quantitative agreement, however, turns out to be rather poor (even within
error bars).

Note that in the vicinity of fairly disordered substrates the nematic becomes
elastically softer than in the bulk, which increases the local slope of the director
profile $\phi(z)$ in comparison with the bulk. As the NI transition is approached,
the thickness of this disordered layer ($\sim\xi$) starts to grow, thereby indirectly
affecting the bulk slope of $\phi(z)$, which --- in principle --- could result in a
more pronounced temperature dependence for $b$ that is extrapolated from the bulk. Note,
however, that a significantly thicker sample than the present one is required to
actually observe this scenario.

\begin{figure}
 \begin{center}
 \includegraphics[width=75mm]{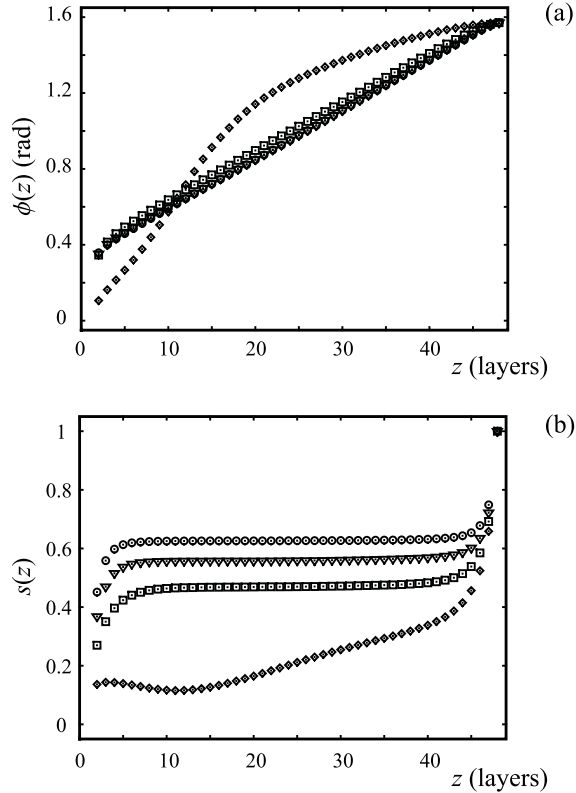}
 \caption[]{Nematic slab with a single free interface at $z=0$ and $\nu=0.05$
            (homeotropic intrinsic anchoring). Same profiles as in Fig.~\ref{sl2},
            yet for rescaled values of $T^*$: $T^*=1.335$, $1.265$, $1.195$, and $1.125$
            (diamonds, squares, triangles, and circles, respectively). $T_{NI}^*\approx 1.36$.}
 \label{sl4}
 \end{center}
\end{figure}

We also studied the influence of the hexagonal lattice structure and the confinement
of ${\mathbf u}_i$ to the $xz$ plane on our results, by analyzing external anchoring in the
original LL model where the spins ${\mathbf u}_i$ are three--dimensional vectors. Both
the LL and hexagonal lattice models yield the same qualitative behavior: a nearly
temperature-independent $b$ which increases in value as the surface roughness is increased.

\begin{figure}
 \begin{center}
 \includegraphics[width=75mm]{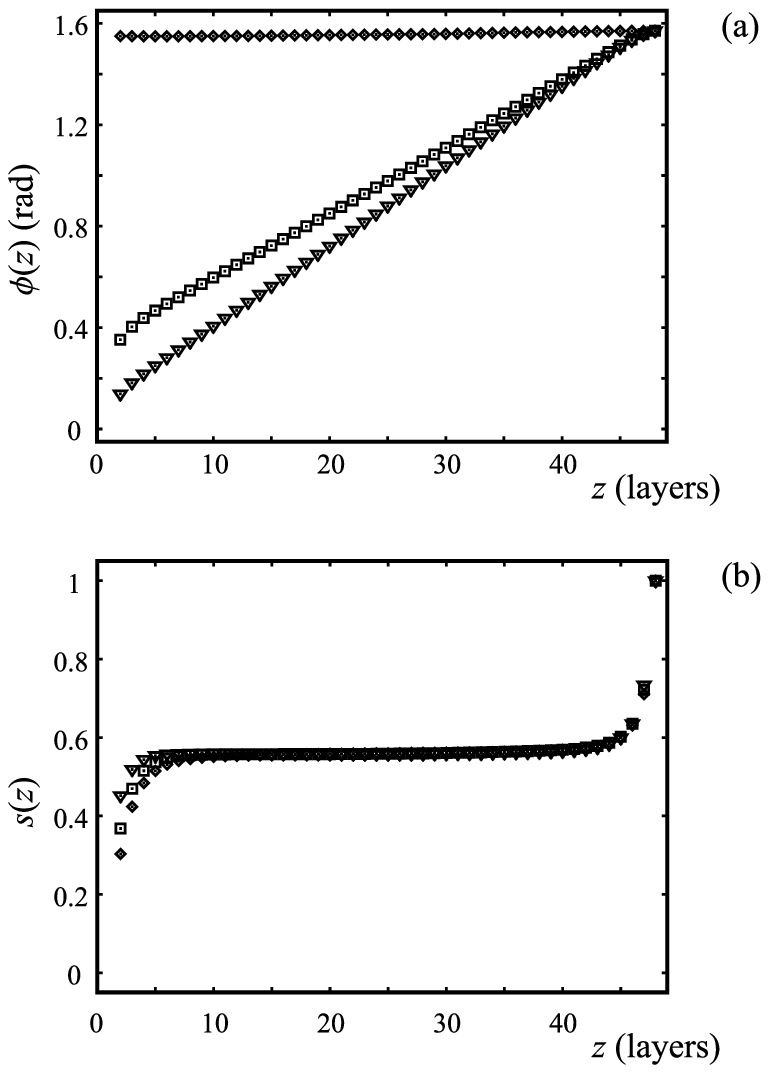}
 \caption[]{Same as Fig.~\ref{sl4}, but for different values of the bulk interaction
            anisotropy: $\nu=0$, $\nu=0.05$, and $\nu=0.1$ (diamonds, squares
            and triangles, respectively). The corresponding temperatures are $T^*=1.34$,
            $T^*=1.195$, and $T^*=1.065$, respectively, ensuring that the bulk value
            of the order parameter and the reduced temperature be the same in all cases
            ($\tau\approx 0.12$). There is no intrinsic anchoring for $\nu=0$, while for
            $\nu\neq 0$ an increase of $\nu$ results in a decrease of the extrapolation
            length.}
 \label{sl5}
 \end{center}
\end{figure}

\section{Intrinsic anchoring}

Setting $\nu\neq 0$, intrinsic anchoring appears at the interfaces, in addition to the
external contribution present already for $\nu=0$. Note that changing $\nu$ affects not
only anchoring, but also changes the elastic softness of the nematic~\cite{Barbero2}: for
$\nu$ not too large ($\nu\lesssim 0.78$), the Frank elastic constant decreases with increasing
$\nu$. Consequently, upon increasing $\nu$, the NI transition temperature is found to decrease
(see Table~\ref{tabela}). Therefore, when comparing $b$ for different values of $\nu$, it is
appropriate to rescale the temperature $T^*$ and compare measurements with comparable
values of the reduced temperature $\tau$ (resulting in similar values of $s_b$).

To facilitate the analysis, we decided to remove the wall at $z=0$ allowing us to deal with intrinsic
anchoring alone. When $\nu\lesssim 0.3$ (but nonzero) the intrinsic easy axis remains homeotropic~\cite{hex},
and the hybrid cell-like geometry studied in the previous section is maintained. As in the case
of pure external anchoring, we find that the intrinsic extrapolation length $b$ shows no temperature
dependence (within estimated error; see Fig.~\ref{sl4}). However, $b$ does depend on the anisotropy of
the nematic-nematic interparticle potential: for $\nu=0.05$ and $\nu=0.1$ one finds $b=16.3a(1\pm 13\%)$
and $b=4.6a(1\pm 10\%)$, respectively (see Table~\ref{tabela} and Fig.~\ref{sl5}). This trend can
be attributed to an increase of the anchoring energy $W$, as well as to a simultaneous decrease
of the elastic constant $K$ upon increasing $\nu$. Note that in the analysis of Ref.~\cite{hex}
$b=8a(1\pm 13\%)$ was found for $\nu=0.1$ at zero temperature. The disagreement with the current
largely temperature--independent estimate can be attributed to considering only nearest neighbors
in the present analysis, which --- as already stated --- results in an underestimation of $b$.

Note, moreover, that for $\nu>0.1$ extrapolation lengths are in the microscopic range (of the
order of a few --- up to 5 --- lattice spacings $a$). On the other hand, experimental values
of the extrapolation length are typically of the order of 100~nm or greater ~\cite{Blinov}.
We can obtain quantitative agreement between our results and experiments by a significant
decrease of the $\nu$ parameter, as also suggested in Ref.~\cite{hex}. A small value of $\nu$ in
Eq.~(\ref{1}) promotes parallel alignment, as is favored, e.g., by steric repulsions in
a system of hard rods. A decrease in $\nu$ might therefore be regarded as an effective
inclusion of steric repulsions which are absent in our lattice model.

\section{Conclusions}

In this paper we have studied the anchoring of a nematic liquid crystal to a
solid substrate and to a free interface, using numerical simulations of a
simple hexagonal lattice model of two--dimensional spins interacting through
a spatially anisotropic potential. We focused on the roles of substrate roughness
and of the spatial anisotropy of the interparticle potential. An elastic deformation
was imposed on the simulation cell, allowing us to extract the surface extrapolation
length $b$ from the director profile. Setting the anisotropy of the potential
to zero allowed us to probe the strength of the external anchoring (the
anchoring arising from the direct interaction of the nematic with the surface). For
temperatures not too close to the nematic--isotropic transition the extrapolation
length was found to be largely temperature independent, and to grow with increasing
surface roughness (as characterized by the distribution of the local preferred axis
on the surface). This increase is physically reasonable, reflecting the decrease in
the overall anchoring of the nematic as the surface becomes more disordered.
Qualitatively similar behavior was found when we simulated the original
Lebwohl--Lasher model where the spins are three--dimensional. 

With the full anisotropic interaction potential present, intrinsic anchoring
arises due to the incomplete spin--spin interactions at the sample surface.
For a free nematic interface (and external anchoring absent), we extracted
the intrinsic surface extrapolation length. Like the extrapolation length in
the presence of only external anchoring, this length is temperature--independent.
We found that the strength of the intrinsic anchoring grows together with the
elastic softness of the nematic as the interaction potential anisotropy is increased,
leading to smaller values of the extrapolation length. Thus, obtaining agreement
with the experimentally measured values of $b$, on the order of 100 nm, requires
a relatively small value of the interaction anisotropy parameter $\nu$ approaching
$10^{-2}$.

\section*{Acknowledgments}

Computational work in support of this research was performed at the Theoretical
Physics Computing Facility at Brown University. Financial support from the
National Science Foundation under Grant Nos. INT--9815313 and DMR--0131573,
and from the Slovenian Office of Science (Programmes No. P0-0503-1554 and 0524-0106)
is gratefully acknowledged.

\end{document}